\documentclass[titlepage]{article}
\usepackage{amsmath}
\usepackage{amssymb}
\usepackage{eucal}
\usepackage{graphicx}
\usepackage{setspace}
\usepackage{rotating}
\usepackage[a4paper,hmargin={2cm,2cm}]{geometry}
\usepackage{pb-diagram}

\newcounter{countalgorithm}
\newtheorem{algorithm}[countalgorithm]{Algorithm}
\newcounter{countdefinition}
\newtheorem{definition}[countdefinition]{Definition}
\newcounter{countremark}
\newtheorem{remark}[countremark]{Remark}
\newcounter{countexample}
\newtheorem{example}[countexample]{Example}

\begin{document}

\title{Sampling Spatially Correlated Clutter}

\author{Oscar H.\ Bustos\footnote{Facultad de Matem\'{a}tica Astronom\'{\i}a y F\'{\i}sica,
Universidad Nacional de C\'{o}rdoba,
Ing.\ Medina Allende esq.\ Haya de la Torre, 5000 C\'ordoba, Argentina,
Fax: 54-351-4334054, \texttt{\{bustos,flesia\}@mate.uncor.edu}}\and
Ana Georgina Flesia\footnotemark[1]\and
Alejandro C.\ Frery\footnote{Universidade Federal de Alagoas,
Instituto de Computa\c{c}\~ao,
Campus A. C. Simões, BR 104 - Norte,
Km 97, Tabuleiro dos Martins - Macei\' o - AL, CEP 57072-970.
\texttt{acfrery@pesquisador.cnpq.br}}\and
Mar\'\i a Magdalena Lucini\footnote{Universidad Nacional de Nordeste,
Facultad de Ciencias Exactas, Naturales y Agrimensura,
Av. Libertad 5450 - Campus "Deodoro Roca",
(3400) Corrientes,
Tel: +54 (3783) 473931/473932\
\texttt{lucini@exa.unne.edu.ar}}}

\maketitle

\begin{abstract}
Correlated ${\cal G}$ distributions can be used to describe the clutter seen in images obtained with coherent illumination, as is the case of B-scan ultrasound, laser, sonar and synthetic aperture radar (SAR) imagery. These distributions are derived using the square root of the generalized inverse Gaussian distribution for the amplitude backscatter within the multiplicative model.
A two-parameters particular case of the amplitude ${\mathcal G}$ distribution, called ${\mathcal G}_{A}^{0}$, constitutes a modeling improvement with respect to the widespread ${\mathcal K}_{A}$ distribution when fitting urban, forested and deforested areas in remote sensing data. This article deals with the modeling and the simulation of correlated ${\mathcal G}_{A}^{0}$-distributed random fields. It is accomplished by means of the Inverse Transform method, applied to Gaussian random fields with spatial correlation.
The main feature of this approach is its generality, since it allows the introduction of negative correlation values in the resulting process, necessary for the proper explanation of the shadowing effect in many SAR images.\\
\textbf{Keywords:} image modeling, simulation, spatial
correlation, speckle.
\end{abstract}


\section{Introduction}

The demand for exhaustive and controlled clutter measurements in all scenarios would be alleviated if plausible data could be obtained by computer simulation.
Clutter simulation is an important element in the development of target detection algorithms for radar, sonar, ultrasound and laser imaging systems.
Using simulated data, the accuracy of clutter models may be assessed and the performance of target detection algorithms may be quantified with controlled clutter backgrounds.
This article is concerned with the simulation of random clutter having appropriate both first and second order statistical properties.

The use of correlation in clutter models is significant and relevant since the correlation effects within the clutter often dominate system performance.
Models merely based on single-point statistics could, therefore, produce misleading results, and several commonly used forms for clutter statistics fall into this category.

The statistical properties of heterogeneous clutter returned by Synthetic Aperture Radar (SAR) sensors have been largely investigated in the literature.
A theoretical model widely adopted for these images assumes that the value in every pixel is the observation of an uncorrelated stochastic process $Z_{A}$, characterized by single-point (first order) statistics.
A general agreement has been reached that amplitude fields are well explained by the ${\mathcal K}_{A}$ distribution.
Such distribution arises when coherent radiation is scattered by a surface having Gamma-distributed cross-section fluctuations.
Though agricultural fields and woodland are very well fitted by this distribution, it is also known that it fails giving accurate statistical description of extremely heterogeneous data, such as urban areas and forest growing on undulated relief.

As discussed in~\cite{frery96,resenhas99}, another distribution, the ${\mathcal G}_{A}$ law, can be used to describe those extremely heterogeneous regions, with the advantage that it has the ${\mathcal K}_{A}$ distribution as a particular case.
This distribution arises in all coherent imaging applications as a result of the action of multiplicative speckle noise on an underlying square root of a generalized inverse Gaussian distribution.
The main drawback of this general model is that it requires an extra parameter, besides its theoretical complexity.

Nevertheless, it can be seen in~\cite{mejailfreryjacobobustos2001,BustosFreryLucini:Mestimators:2001,CribariFrerySilva:CSDA} that a special case of the ${\mathcal G}_{A}$ distribution, namely the ${\mathcal G}_{A}^{0}$ law, which has as many parameters as the ${\mathcal K}_{A}$ distribution, is able to model with accuracy every type of clutter.
As a consequence, efforts have been directed toward the simulation of ${\mathcal G}_{A}^{0}$ textures, but no exact method for generating patterns with arbitrary spatial autocorrelation functions has been envisaged so far, in spite of it being more tractable than the ${\mathcal K}_{A}$ distribution.

As previously stated, spatial correlation is needed in order to increase the adequacy of the model to real situations.
This paper tackles the problem of simulating correlated ${\mathcal G}_{A}^{0}$ fields.

\section[Correlated GA0 clutter]{Correlated ${\mathcal G}_{A}^{0}$ clutter}

The main properties and definitions of the ${\mathcal G}_{A}^{0}$ clutter are presented in this section, starting with the first order properties of the distribution and concluding with the definition of a ${\mathcal G}_{A}^{0}$ stochastic process that will describe $Z_A$ fields.

\subsection{Marginal properties}

The ${\mathcal G}_{A}^{0}(\alpha,\gamma,n)$ distribution is characterized by the following probability density function:
\begin{equation}
f_{Z_{A}}(z,(\alpha,\gamma,n))=\frac{2n^{n}\Gamma(n-\alpha)}{\sqrt{\gamma
}\Gamma(-\alpha)\Gamma(n)}\cdot\frac{\left(  \frac{z}{\sqrt{\gamma}}\right)
^{2n-1}}{\left(  1+\frac{z^{2}}{\gamma}n\right)^{n-\alpha}}\cdot \mathbb{I}_{(0,+\infty
)}(z),\quad\alpha<0,\gamma>0,\label{uno}
\end{equation}
being $n\geq1$ the number of looks of the image, which is controlled at the
image generation process, and $\mathbb{I}_T(\cdot)$ the indicator function of the set $T$.
The parameter $\alpha$ describes the roughness, being small values (say $\alpha\leq -15$) usually associated to homogeneous targets, like pasture, values ranging in the $(-15,-5]$ interval usually observed in heterogeneous clutter, like forests, and big values ($-5<\alpha<0$ for instance) commonly seen when extremely heterogeneous areas are imaged.
The parameter $\gamma$ is related to the scale, in the sense that if $Z$ is ${\mathcal G}_{A}^{0}(\alpha,1,n)$ distributed then  $Z_{A}=\sqrt{\gamma}Z$ obeys a ${\mathcal G}_{A}^{0}(\alpha,\gamma,n)$ law.

\begin{figure}[h]
\begin{center}
\includegraphics[width=15cm]{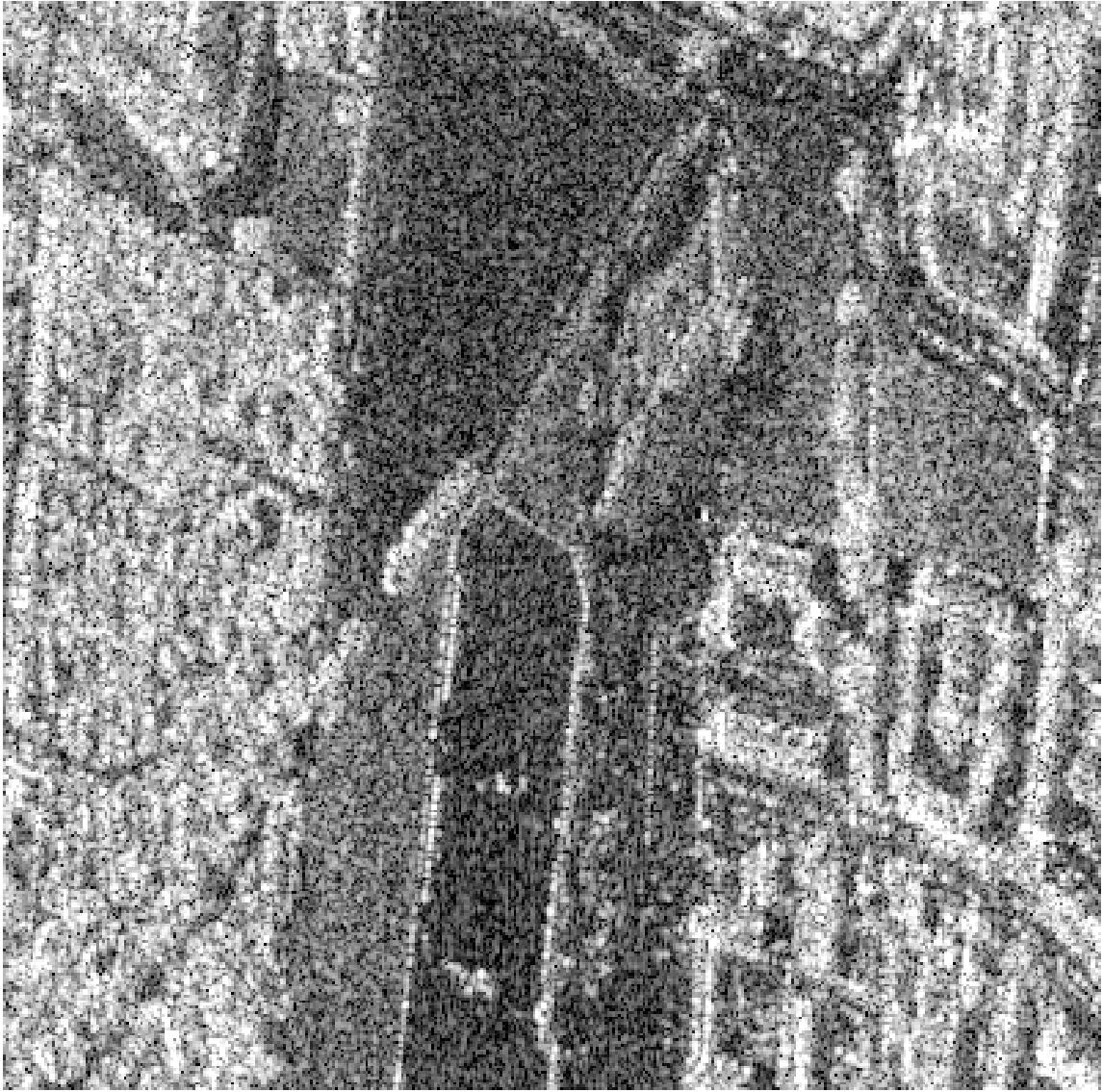}
\caption{E-SAR image showing three types of texture.}
\label{esarimage}
\end{center}
\end{figure}

A SAR image over a suburban area of M\"unchen, Germany, is shown in Figure~\ref{esarimage}.
It was obtained with E-SAR, an experimental polarimetric airborne sensor operated by the German Aerospace Agency (Deutsches Zentrum f\"ur Luft- und Raumfahrt -- DLR e.\ V.)
The data here shown were generated in single look format, and exhibit the three discussed types of roughness: homogeneous (the dark areas to the middle of the image), heterogeneous (the clear area to the left) and extremely heterogeneous (the clear area to the right).

The $r$-th moments of the ${\mathcal G}_{A}^{0}(\alpha,\gamma,n)$ distribution are
\begin{equation}
{E}(Z_{A}^{r})=
\left(\frac{\gamma}{n}\right)
^{\frac{r}{2}}\frac{\Gamma(-\alpha-\frac{r}{2})\Gamma(n+\frac{r}{2})}%
{\Gamma(-\alpha)\Gamma(n)},\qquad\alpha<-r/2,n\geq 1, \label{moments}%
\end{equation}
when $-r/2\leq\alpha<0$ the $r$-th order moment is infinite. Using
equation~(\ref{moments}) the mean and variance of a ${\mathcal G}_{A}^{0}%
(\alpha,\gamma,n)$ distributed random variable can be computed:%
\begin{align*}
\mu_{Z_{A}}  &  =\sqrt{\frac{\gamma}{n}}\frac{\Gamma(n+\frac{1}{2}%
)\Gamma(-\alpha-\frac{1}{2})}{\Gamma(n)\Gamma(-\alpha)},\\
\sigma_{Z_{A}}^{2}  &  =\frac{\gamma\left[  n\Gamma^{2}(n)(-\alpha
-1)\Gamma^{2}(-\alpha-1)-\Gamma^{2}(n+\frac{1}{2})\Gamma^{2}(-\alpha
-\frac{1}{2})\right]  }{n\Gamma^{2}(n)\Gamma^{2}(-\alpha)}.
\end{align*}

\begin{figure}[h]
\begin{center}
\includegraphics[width=15cm]{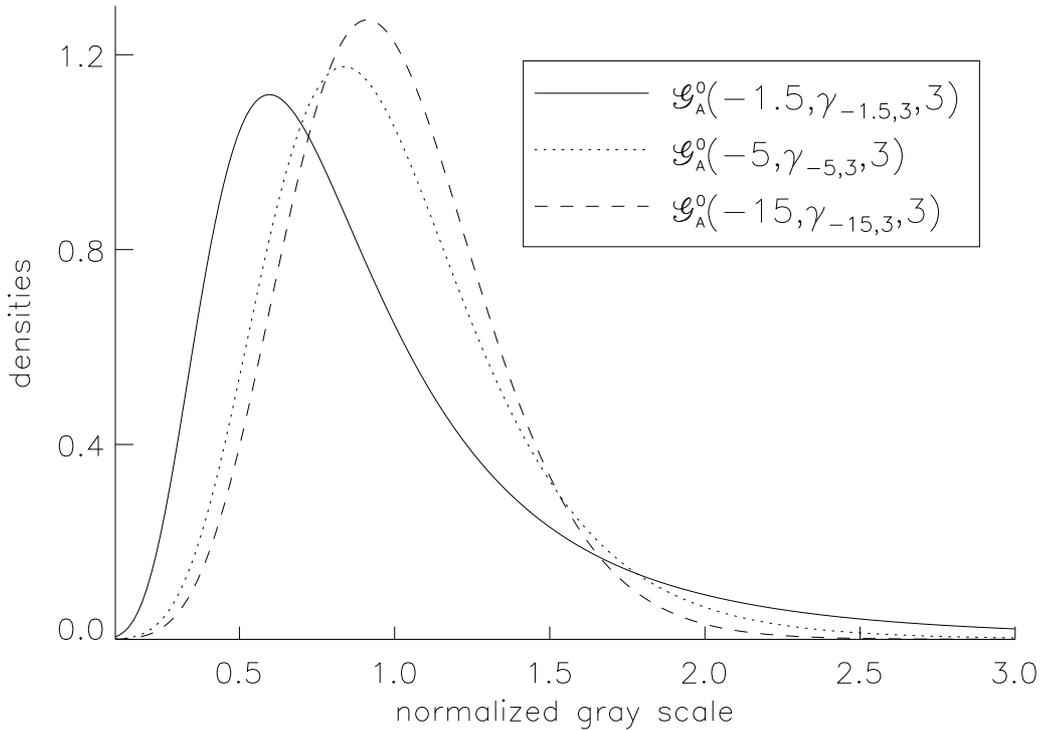}%
\caption{Densities of the ${\mathcal G}_{A}^{0}(\alpha,\gamma_{\alpha,3},3)$
distribution.}
\label{densga0}
\end{center}
\end{figure}

Figure~\ref{densga0} shows three densities of the ${\mathcal G}_{A}^{0}(\alpha,\gamma,n)$ distribution for the single look ($n=1$) case. These densities are normalized so that the expected value is $1$ for every value of the roughness parameter.
This is obtained using equation~(\ref{moments}) for setting the scale parameter $\gamma=\gamma_{\alpha,n}=n\left(  \Gamma
(-a)\Gamma(n)/\left(  \Gamma(-a-1/2)\Gamma(n+1/2)\right)  \right)  ^{2}$.
These densities illustrate the three typical situations described above: homogeneous areas ($\alpha=-15$, dashes), heterogeneous clutter ($\alpha=-5$, dots) and an extremely heterogeneous target ($\alpha=-1.5$, solid line).

\begin{figure}[h]
\begin{center}
\includegraphics[width=15cm]{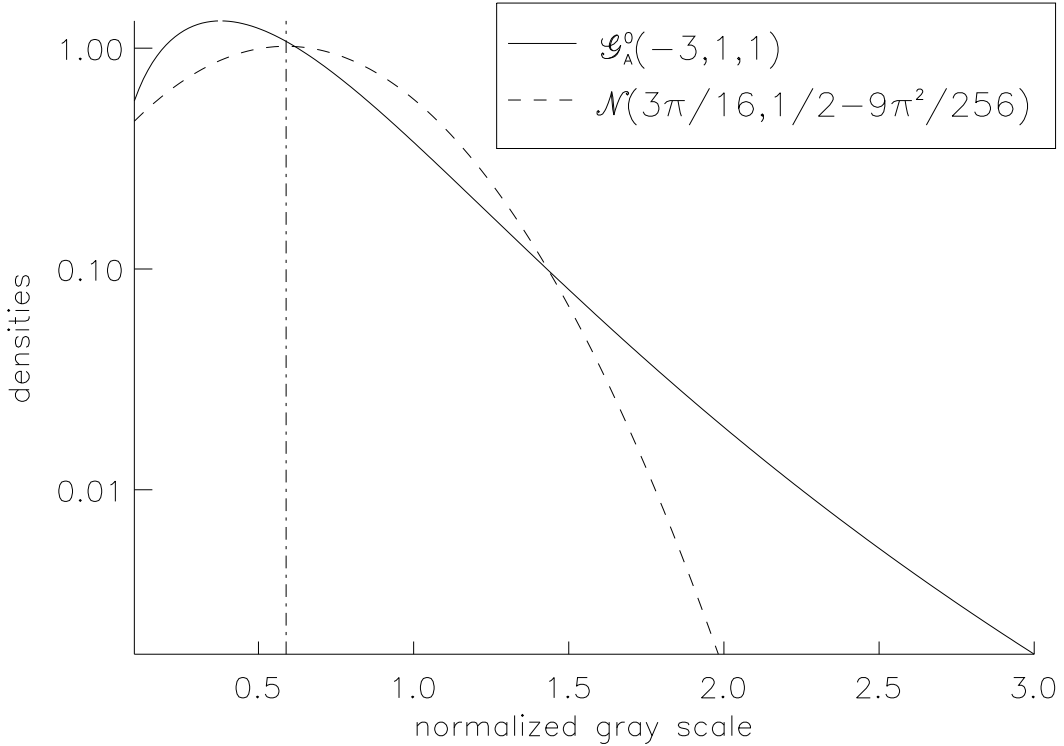}
\caption{Densities of the ${\mathcal G}_{A}^{0}$ and
Gaussian distributions with same mean values $\mu=3\pi/16$ in
semilogarithmic scale.}
\label{logdens}
\end{center}
\end{figure}

Following Barndorff-Nielsen and Bl\ae sild~\cite{barndorffblaesild81}, it is interesting to see these densities as log probability functions, particularly because the ${\mathcal G}_{A}^{0}$ is closely related to the class of Hyperbolic distributions~\cite{florence}.
Figure~\ref{logdens} shows the densities of the ${\mathcal G}_{A}^{0}(-3,1,1)$ and ${\mathcal N}(3\pi
/16,1/2-9\pi^{2}/256)$ distributions in semilogarithmic scale, along with their mean value $\mu=3\pi/16$. The parameters were chosen so that these distributions have equal mean and variance.
The different decays of their tails is evident: the former behaves logarithmically, while the latter decays quadratically.
This behavior ensures the ability of the ${\mathcal G}_{A}^{0}$ distribution to model data with extreme variability.

Besides being essential for the simulation technique here proposed, cumulative distribution functions are needed for carrying out goodness of fit tests and for the proposal of estimators based on order statistics.
It can be seen in~\cite{mejailfreryjacobobustos2001,MejailJacoboFreryBustos:IJRS,martajuliofreryoscar} that the cumulative distribution function of a ${\mathcal G}_{A}^{0}(\alpha,\gamma,n)$ distributed random variable is given, for every $z>0$, by $G(z,(\alpha,\gamma,n))=\Upsilon_{2n,-2\alpha
}(-\alpha z^{2}/\gamma)$, where $\Upsilon_{s,t}$ is the cumulative
distribution function of a Snedecor's $F_{s,t}$ distributed random variable
with $s$ and $t$ degrees of freedom.
Both $\Upsilon_{\cdot,\cdot}$ and $\Upsilon^{-1}_{\cdot,\cdot}$ are readily available in most platforms for computational statistics.

The single look case is of particular interest since it describes the noisiest images and it exhibits nice analytical properties.
The distribution is characterized by the density $f(z;\alpha ,\gamma, 1 )=-\frac{2\alpha }{\gamma ^{\alpha }}\,z(\gamma
+z^{2})^{\alpha -1}\mathbb{I}_{(0,\infty)}(z)$, whith $-\alpha ,\gamma >0$. Its cumulative distribution function is given by $F(t)=1-\left( 1+t^{2}/\gamma \right) ^{\alpha }\mathbb{I}_{(0,\infty)}(t)$, and its inverse, useful for the generation of random deviates and the computation of quantiles, is given by $F^{-1}(t)=\left(\gamma \left( (1-t)^{1/\alpha }-1\right)\right)^{1/2}\mathbb{I}_{(0,1)}(t)$.

\subsection{Correlated clutter}

Instead of defining the model over $\mathbb{Z}^2$, in this section a realistic description of finite-sized fields is made.
Let $Z_{A}=(Z_{A}(k,\ell))_{0\leq k\leq N-1,0\leq\ell\leq N-1}$ be the
stochastic model that describes the return amplitude image.

\begin{definition}
We say that $Z_{A}$ is a ${\mathcal G}_{A}^{0}(\alpha,\gamma,n)$ stochastic
process with correlation function $\rho_{Z_{A}}$ (in symbols $Z_{A}$
$\sim({\mathcal G}_{A}^{0}(\alpha,\gamma,n),\rho_{Z_{A}})$) if for all $0\leq
i,j,k,\ell\leq N-1$ holds that
\begin{enumerate}
\item $Z_{A}(k,\ell)$ obeys a ${\mathcal G}_{A}^{0}(\alpha,\gamma,n)$ law;
\item the mean field is $\mu_{Z_{A}}=E(Z_{A}(k,\ell))$;
\item the variance field is $\sigma_{Z_{A}}^{2}=Var(Z_{A}(k,\ell))$;
\item the correlation function is $\rho_{Z_{A}}((i,j),(k,\ell))=\left(
E(Z_{A}(i,j)Z_{A}(k,\ell))-\mu_{Z_{A}}^{2}\right)  /\sigma_{Z_{A}}^{2}$.
\end{enumerate}
\end{definition}

The scale property of the parameter $\gamma$ implies that correlation function $\rho_{Z_{A}}$ and $\gamma$ are unrelated and, therefore, it is enough to generate a $Z_{A}^{1}\sim({\mathcal G}_{A}^{0}(\alpha,1,n),\rho_{Z_{A}})$ field and then simply multiply every outcome by $\gamma^{1/2}$ to get the desired field.

This paper presents a variation of a method used for simulation of correlated Gamma variables, called Transformation Method, that can be found
in~\cite{BustosFlesiaFrery99}.
This method can be summarized in the following three steps:
\begin{enumerate}
\item Generate independent outcomes from a convenient distribution.
\item \label{step2}Introduce correlation in these data.
\item Transform the correlated observations into data with the desired
marginal properties~\cite{oliverquegan98}.
\end{enumerate}

The transformation that guarantees the validity of this procedure is obtained from the cumulative distribution functions of the data obtained in
step~\ref{step2}, and from the desired set of distributions.

Recall that if $U$ is a continuous random variable with cumulative distribution function $F_{U}$ then $F_{U}(U)$ obeys a uniform ${\mathcal U}(0,1)$ law and, reciprocally, if $V$ obeys a ${\mathcal U}(0,1)$
distribution then $F_{U}^{-1}(V)$ is $F_{U}$ distributed.
In order to use this method it is necessary to know the correlation that the random variables will have after the transformation, besides the function $F_{U}^{-1}$.

The method here studied consists of the following steps:
\begin{enumerate}
\item propose a correlation structure for the ${\mathcal G}_{A}^{0}$ field,
say, the function $\rho_{Z_{A}}$;
\item generate a field of independent identically distributed standard
Gaussian observations;
\item compute $\tau$, the correlation structure to be imposed to the
Gaussian field from $\rho_{Z_{A}}$, and impair it using the Fourier
transform without altering the marginal properties;
\item transform the correlated Gaussian field into a field of observations
of identically distributed ${\mathcal U}(0,1)$ random variables, using the
cumulative distribution function of the Gaussian distribution ($\Phi$);
\item transform the uniform observations into ${\mathcal G}_{A}^{0}$
outcomes, using the inverse of the cumulative distribution function of the
${\mathcal G}_{A}^{0}$ distribution ($G^{-1}$).
\end{enumerate}

The function that relates $\rho_{Z_{A}}$ and $\tau$ is computed using
numerical tools.
In principle, there are no restrictions on the possible roughness parameters values that can be obtained by this method, but issues
related to machine precision must be taken into account.
Another important issue is that not every desired final correlation structure $\rho_{Z_{A}}$ is mapped onto a feasible intermediate correlation structure $\tau$.
The procedure is presented in detail in the next section.

\section{Transformation Method}

Let $G(\cdot,(\alpha,\gamma,n))$ be the cumulative distribution function of a ${\mathcal G}_{A}^{0}(\alpha,\gamma,n)$ distributed random variable. As previously stated,
\[
G(x,(\alpha,\gamma,n))={\Upsilon}_{2n,-2\alpha}\left(
-\frac{\alpha x^{2}}{\gamma}\right)  ,
\]
where $\Upsilon_{\nu_{1},\nu_{2}}$ is the cumulative distribution function of a Snedecor $F_{_{\nu_{1},\nu_{2}}}$ distribution, i.e.,%
\[
{\Upsilon}_{\nu_{1},\nu_{2}}(x)=\frac{\Gamma\left(  \frac{\nu
_{1}+\nu_{2}}{2}\right)  }{\Gamma\left(  \frac{\nu_{1}}{2}\right)
\Gamma\left(  \frac{\nu_{2}}{2}\right)  }\left(  \frac{\nu_{1}}{\nu_{2}%
}\right)  ^{\frac{\nu_{1}}{2}}\int_{0}^{x}t^{\frac{\nu_{1}-2}{2}}\left(
1+\frac{\nu_{1}}{\nu_{2}}t\right)  ^{-\frac{\nu_{1}+\nu_{2}}{2}}dt.
\]

The inverse of $G(\cdot,(\alpha,\gamma,n))$ is, therefore,
\[
G^{-1}(t,(\alpha,\gamma,n))=\sqrt{-\frac{\gamma}{\alpha}{\Upsilon
}_{2n,-2\alpha}^{-1}(t)}.
\]

To generate $Z_{A}^{1}=(Z_{A}^{1}(k,\ell))_{0\leq k\leq N-1,0\leq\ell\leq
N-1}\sim({\mathcal G}_{A}^{0}(\alpha,1,n),\rho_{Z_{A}})$ using the inversion method we define every coordinate of the process $Z_A$ as a transformation of a Gaussian process $\zeta$ as $ Z_{A}^{1}(i,j)=$ $G^{-1}(\Phi(\zeta(i,j)),(\alpha,1,n))$, where $\zeta
=(\zeta(i,j))_{0\leq i\leq N-1,0\leq j\leq N-1}$ is a stochastic process such that $\zeta(i,j)$ is a standard Gaussian random variable and with correlation function $\tau_{\zeta}$ (i.e. where $\tau_{\zeta}((i,j),(k,\ell))=E(\zeta (i,j)\zeta(k,\ell))$) satisfying
\begin{equation}
\rho_{Z_{A}}((i,j),(k,\ell))=\varrho_{(\alpha,n)}(\tau_{\zeta}((i,j),(k,\ell)))
\label{dos}%
\end{equation}
for all $0\leq i,j,k,\ell\leq N-1$ and $(i,j)\neq(k,\ell)$ and where $\Phi$
denotes the cumulative distribution function of a standard Gaussian random
variable.

Posed as a diagram, the method consists of the following transformations
among Gaussian ($\mathcal N$), Uniform ($\mathcal U$) and ${\mathcal G}_A^0$-distributed random variables:
\[
\begin{diagram}
\node{\mathcal{N}}\arrow{e,t}{\Phi}
\node{\mathcal{U}}\arrow{s,r}{G^{-1}}\\
\node{}\node{\mathcal{G}_A^0}
\end{diagram}
\]

A central issue of the method is finding the correlation structure that the Gaussian field has to obey, in order to have the desired $\mathcal{G}_A^0$ field after the transformation.
The function $\varrho_{(\alpha,n)}$ is defined on $(-1,1)$ by
\[
\varrho_{(\alpha,n)}(\tau)=\frac{R_{(\alpha,n)}(\tau)-\left(  \frac{1}{n}\right)
\left(  \frac{\Gamma(n+\frac{1}{2})\Gamma(-\alpha-\frac{1}{2})}{\Gamma
(n)\Gamma(-\alpha)}\right)  ^{2}}{-\frac{1}{1+\alpha}-\left(  \frac{1}%
{n}\right)  \left(  \frac{\Gamma(n+\frac{1}{2})\Gamma(-\alpha-\frac{1}{2}%
)}{\Gamma(n)\Gamma(-\alpha)}\right)  ^{2}},
\]
with
\begin{align*}
R_{(\alpha,n)}(\tau)  &  = \iint_{\mathbb{R}^2}
G^{-1}(\Phi(u),(\alpha,1,n))G^{-1}(\Phi(v),(\alpha,1,n))\phi_{2}%
(u,v,\tau)))dudv\\
&  =\frac{1}{\left|  \alpha\right|  2\pi\sqrt{1-\tau^{2}}}
\iint_{\mathbb{R}^2}
\sqrt{{ \Upsilon}_{2n,-2\alpha
}^{-1}(\Phi(u)).{\Upsilon}_{2n,-2\alpha}^{-1}(\Phi(v))} \exp\left(
-\frac{u^{2}-2\tau.u.v+v^{2}}{2(1-\tau^{2})}\right)  dudv,
\end{align*}
where%
\[
\phi_{2}(u,v,\tau)=\frac{1}{2\pi\sqrt{(1-\tau^{2})}}\exp\left(  -\frac{u^{2}%
-2\tau.u.v+v^{2}}{2(1-\tau^{2})}\right)  .
\]
Note that $R_{(\alpha,n)}(\tau_{\zeta}((i,j),(k,\ell)))=E(Z_{A}^{1}%
(i,j)Z_{A}^{1}(k,\ell))$ for all $0\leq i,j,k,\ell\leq N-1$ and $(i,j)\neq
(k,\ell)$.

The answer to the question of finding $\tau_{\zeta}$ given $\rho
_{\mathbf{Z_{A}}}$ is equivalent to the problem of inverting the function
$\varrho_{(\alpha,n)}$. This function is only available using numerical methods, an
approximation that may impose restrictions on the use of this simulation method.

\subsection[Inversion of the correlation function]{Inversion of $\varrho_{(\alpha,n)}$}

The function $\varrho_{(\alpha,n)}$ has the following properties:
\begin{enumerate}
\item The set $\{\varrho_{(\alpha,n)}(\tau)\colon\tau\in(-1,1)\}$ is strictly included
in $(-1,1)$, and depends on the values of $\alpha$.
\item The function $\varrho_{(\alpha,n)}$ is strictly increasing in $(-1,1)$.
\item The values $\varrho_{(\alpha,n)}(\tau)$ are strictly negative for all
$\tau<0$.
\end{enumerate}

Let $\eth_{(\alpha,n)}$ be the inverse function of $\varrho_{(\alpha,n)}$. Then, in
order to calculate its value for a fixed $\rho\in(-1,1)$, we have to solve the
following equation in $\tau$:%
\[
R_{(\alpha,n)}(\tau)+\frac{\rho}{1+\alpha}+\left(  \rho-1\right)  \left(
\frac{1}{n}\right)  \left(  \frac{\Gamma(n+\frac{1}{2})\Gamma(-\alpha
-\frac{1}{2})}{\Gamma(n)\Gamma(-\alpha)}\right)  ^{2}=0
\]

\begin{figure}[h]
\begin{center}
\includegraphics[width=15cm]{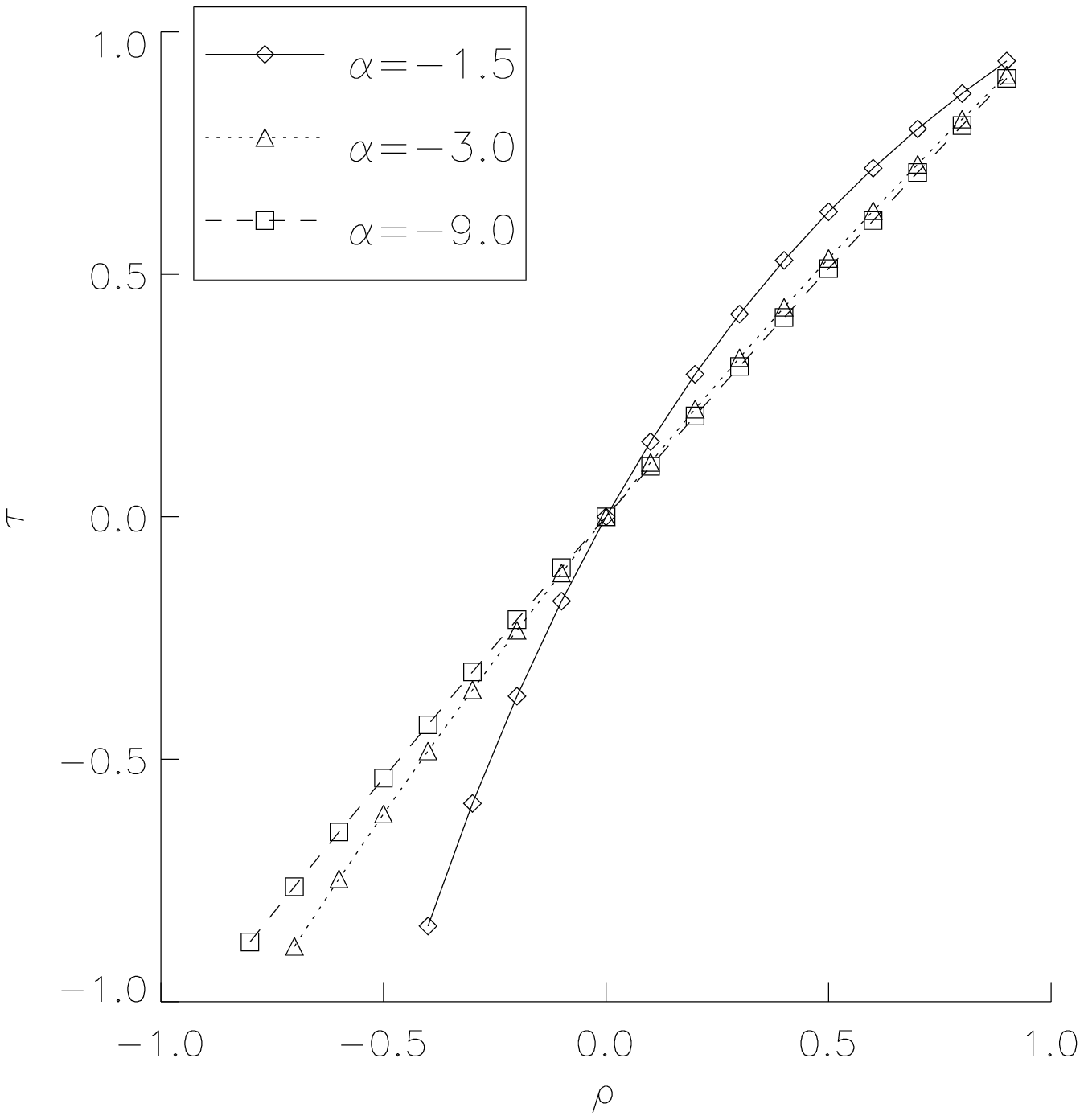}
\caption{Values of $\tau$ as a function of $\rho$ for $n=1$ and varying $\alpha$.}
\label{figtau1}
\end{center}
\end{figure}

Then, it follows from the properties of $\varrho_{(\alpha,n)}$, that for certain values of $\alpha$ the set of $\tau$ such that this equation is solvable is a strict subset of $(-1,1)$. Table~\ref{tabla1} shows some values of the function $\eth_{(\alpha,n)}$ for specific values of $\rho$, $n$ and $\alpha$.
Figure~\ref{figtau1} shows $\tau$ as a function of $\rho$ for the $n=1$ case and varying values of $\alpha$, and it can be seen that the smaller $\alpha$ the closer this function is to the identity.
This is sensible, since the ${\mathcal G}_A^0$ distribution becomes more and more symmetric as $\alpha\to-\infty$ and, therefore, simulating outcomes from this distribution becomes closer and closer to the problem of obtaining Gaussian deviates.

\begin{figure}[h]
\begin{center}
\includegraphics[width=15cm]{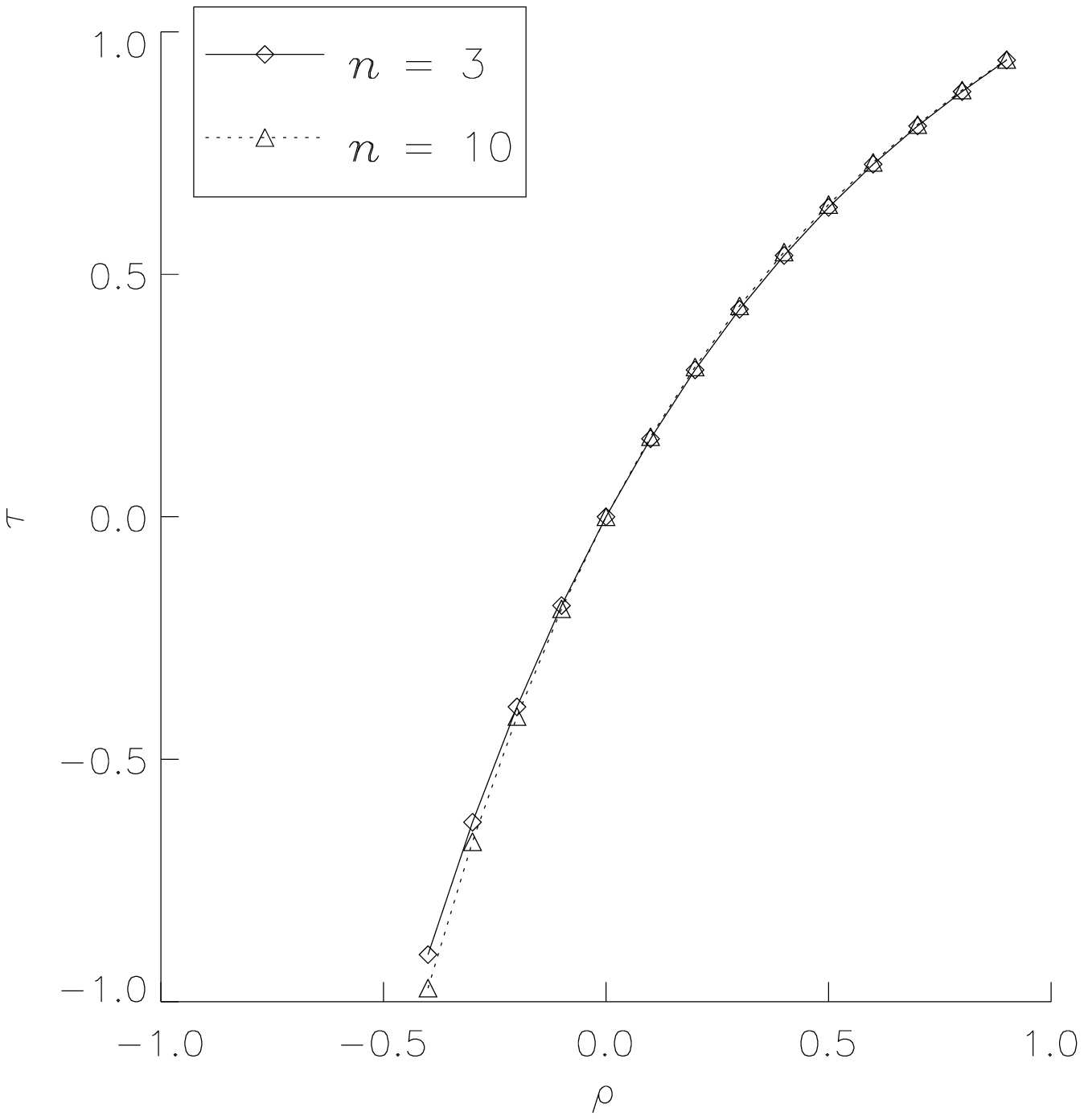}
\caption{Values of $\tau$ as a function of $\rho$ for $\alpha=-1.5$ and varying $n$.}
\label{figtau2}%
\end{center}
\end{figure}

Figure~\ref{figtau2} presents the same function for $\alpha=-1.5$ and varying number of looks.
It is noticeable that $\tau$ is far less sensitive to $n$ than to $\alpha$, a feature that suggests a shortcut for computing the values of Table~\ref{tabla1}: disregarding the dependence on $n$, i.e., considering $\tau(\rho,\alpha,n)\simeq\tau (\rho,\alpha,n_{0})$ for a fixed convenient $n_{0}$.

\begin{table}{h}
\begin{tabular}[c]{|c||c|c|c|c||c|c|c|c||c|c|c|c|}\hline
\multicolumn{13}{|c|}{}\\ \hline\hline
$\rho$ & \multicolumn{4}{||c||}{$\alpha=-1.5$} &
\multicolumn{4}{||c||}{$\alpha=-3.0$} & \multicolumn{4}{||c|}{$\alpha=-9.0$%
}\\\cline{2-13}
& $n=1$ & $n=3$ & $n=6$ & $n=10$ & $n=1$ & $n=3$ & $n=6$ & $n=10$ & $n=1$ &
$n=3$ & $n=6$ & $n=10$\\\hline\hline
$-.9$ &  &  &  &  &  &  &  &  &  & $-.953$ & $-.954$ & $-.958$\\\hline
$-.8$ &  &  &  &  &  &  &  &  & $-.877$ & $-.845$ & $-.845$ & $-.848$\\\hline
$-.7$ &  &  &  &  & $-.886$ & $-.881$ & $-.901$ & $-.915$ & $-.763$ & $-.737$
& $-.737$ & $-.740$\\\hline
$-.6$ &  &  &  &  & $-.747$ & $-.745$ & $-.761$ & $-.772$ & $-.650$ & $-.630$
& $-.630$ & $-.632$\\\hline
$-.5$ &  &  &  &  & $-.613$ & $-.612$ & $-.624$ & $-.632$ & $-.539$ & $-.523$
& $-.523$ & $-.525$\\\hline
$-.4$ & $-.844$ & $-.903$ & $-$.$948$ & $-.972$ & $-.483$ & $-.483$ & $-.492$
& $-.498$ & $-.429$ & $-.417$ & $-.417$ & $-.419$\\\hline
$-.3$ & \ $-.591$ & $-.630$ & $-$.$656$ & $-.670$ & $-.357$ & $-.357$ &
$-.363$ & $-.367$ & $-.320$ & $-.312$ & $-.312$ & $-.313$\\\hline
$-.2$ & $-.370$ & $-.392$ & $-$.$405$ & $-.412$ & $-.234$ & $-.235$ & $-.239$
& $-.241$ & $-.212$ & $-.207$ & $-.207$ & $-.208$\\\hline
$-.1$ & $-.174$ & $-.183$ & $-$.$188$ & $-.190$ & $-.116$ & $-.116$ & $-.117$
& $-.119$ & $-.105$ & $-.103$ & $-.103$ & $-.104$\\\hline
$0$ & $.0$ & $.0$ & .$0$ & $.0$ & $.0$ & $.0$ & $.0$ & $.0$ & $.0$ & $.0$ &
$.0$ & $.0$\\\hline
$.1$ & $.155$ & $.161$ & .$164$ & $.165$ & $.112$ & $.113$ & $.114$ & $.115$ &
$.104$ & $.103$ & $.103$ & $.103$\\\hline
$.2$ & $.294$ & $.303$ & .$307$ & $.309$ & $.222$ & $.223$ & $.225$ & $.226$ &
$.208$ & $.205$ & $.205$ & $.205$\\\hline
$.3$ & $.418$ & $.428$ & .$433$ & $.435$ & $.328$ & $.329$ & $.332$ & $.334$ &
$.310$ & $.306$ & $.306$ & $.307$\\\hline
$.4$ & $.529$ & $.539$ & .$544$ & $.546$ & $.432$ & $.433$ & $.436$ & $.438$ &
$.411$ & $.407$ & $.407$ & $.408$\\\hline
$.5$ & $.629$ & $.638$ & .$642$ & $.644$ & $.533$ & $.534$ & $.537$ & $.539$ &
$.512$ & $.507$ & $.508$ & $.508$\\\hline
$.6$ & $.719$ & $.727$ & .$730$ & $.731$ & $.631$ & $.633$ & $.635$ & $.637$ &
$.611$ & $.607$ & $.607$ & $.608$\\\hline
$.7$ & $.800$ & $.806$ & .$808$ & $.809$ & $.727$ & $.728$ & $.731$ & $.732$ &
$.710$ & $.706$ & $.706$ & $.707$\\\hline
$.8$ & $.873$ & $.877$ & .$879$ & $.880$ & $.820$ & $.821$ & $.823$ & $.824$ &
$.807$ & $.805$ & $.805$ & $.805$\\\hline
$.9$ & $.940$ & $.942$ & .$942$ & $.943$ & $.911$ & $.912$ & $.913$ & $.913$ &
$.904$ & $.903$ & $.903$ & $.903$\\\hline
\end{tabular}
\caption{Values of function $\eth_{(\alpha,n)}$.}
\label{tabla1}
\end{table}

The source FORTRAN file with routines for computing the functions
$\varrho_{(\alpha,n)}$ and $\eth_{(\alpha,n)}$ can be obtained from the first author of this paper.

\subsection[Generation of the Gaussian Process]{Generation of the process $\zeta$}

The process $\zeta$, that consists of spatially correlated standard Gaussian random variables, will be generated using a spectral technique that employs the Fourier transform.
This method has computational advantages with respect to the direct application of a convolution filter.
Again, the concern here is to define a finite process instead of working on $\mathbb{Z}^2$ for the sake of simplicity.

Consider the following sets:
\begin{align*}
R_{1}  &  =\{(k,\ell)\colon 0\leq k,\ell\leq N/2\},\\
R_{2}  &  =\{(k,\ell)\colon N/2+1\leq k\leq N-1,0\leq\ell\leq N/2\},\\
R_{3}  &  =\{(k,\ell)\colon 0\leq k\leq N/2,N/2+1\leq\ell\leq N-1\},\\
R_{4}  &  =\{(k,\ell)\colon N/2+1\leq k\leq N-1,N/2+1\leq\ell\leq N-1\},\\
R_{N}  &  =R_{1}\cup R_{2}\cup R_{3}\cup R_{4}=\{(k,\ell)\colon 0\leq k,\ell\leq
N-1\},\\
\overline{R_{N}}  &  =\{(k,\ell)\colon -(N-1)\leq k,\ell\leq N-1\}.
\end{align*}

Let $\rho\colon R_{1}\longrightarrow(-1,1)$ be a function, extended onto $\overline
{R_{N}}$ by:
\[
\rho(k,\ell)=\left\{
\begin{array}
[c]{ccc}%
\rho(N-k,\ell) & \mathrm{if} & (k,\ell)\in R_{2},\\
\rho(k,N-\ell) & \mathrm{if} & (k,\ell)\in R_{3},\\
\rho(N-k,N-\ell) & \mathrm{if} & (k,\ell)\in R_{4},\\
\rho(N+k,\ell) & \mathrm{if} & -(N-1)\leq k<0\leq\ell\leq N-1,\\
\rho(k,N+\ell) & \mathrm{if} & -(N-1)\leq\ell<0\leq k\leq N-1,\\
\rho(N+k,N+\ell) & \mathrm{if} & -(N-1)\leq k,\ell<0.
\end{array}
\right.
\]

Let $Z_{A}=(Z_{A}(k,\ell))_{0\leq k\leq N-1,0\leq l\leq N-1}$ be a
${\mathcal G}_{A}^{0}(\alpha,\gamma,n)$ stochastic process with correlation
function $\rho_{Z_{A}}$ defined by
\[
\rho_{Z_{A}}((k_{1},\ell_{1}),(k_{2},\ell_{2}))=\rho(k_{2}-k_{1},\ell_{2}%
-\ell_{1}).
\]
Assume that $\tau(k,\ell)=\eth_{(\alpha,n)}(\rho(k,\ell))$ is defined for all $(k,\ell)$ in $R_{N}$.

Let ${\mathcal F}(\tau)\colon R_{N}\longrightarrow\mathbb{C}$ be the normalized Fourier Transform of $\tau$, that is,
\[
{\mathcal F}(\tau)(k,\ell)=\frac{1}{N^{2}}\sum_{k_{1}=0}^{N-1}\sum_{\ell_{1}%
=0}^{N-1}\tau(k_{1},\ell_{1})\exp(-2\pi i(k\cdot k_{1}+\ell\cdot\ell_{1})/N^{2}).
\]
Let $\psi\colon R_{N}\longrightarrow\mathbb{C}$ be defined by $\psi(k,\ell
)=\sqrt{{\mathcal F}(\tau)(k,\ell)}$ and let the function $\theta
:\overline{R_{N}}=\{(k,\ell)\colon -(N-1)\leq k,\ell\leq N-1\}\longrightarrow\mathbb{R}$
be defined by
\[
\theta(k,\ell)={\mathcal F}^{-1}(\psi)(k,\ell)/N=\frac{1}{N}\sum_{k_{1}%
=0}^{N-1}\sum_{\ell_{1}=0}^{N-1}\psi(k_{1},\ell_{1})\exp(2\pi i(k\cdot k_{1} +\ell\cdot \ell_{1})/N^{2}),
\]
(the normalized inverse Fourier Transform of $\psi$) for all $(k,\ell
)\in{R_{N}}$; and
\[
\theta(k,\ell)=\left\{
\begin{array}
[c]{ccc}%
\theta(N+k,\ell) & \mathrm{if} & -(N-1)\leq k<0\leq\ell\leq N-1,\\
\theta(k,N+\ell) & \mathrm{if} & -(N-1)\leq\ell<0\leq k\leq N-1,\\
\theta(N+k,N+\ell) & \mathrm{if} & -(N-1)\leq k,\ell<0.
\end{array}
\right.
\]
A straightforward calculation shows that%
\[
(\theta\ast\theta)(k,\ell)=\sum_{k_{1}=0}^{N-1}\sum_{\ell_{1}=0}^{N-1}%
\theta(k_{1},\ell_{1})\theta(k-k_{1},\ell-\ell_{1})=\tau(k,\ell),
\]
for all $(k,\ell)\in R_{N}$.

\begin{remark}
We can see that ${\mathcal F}(\tau)(k,\ell)\geq0$ and the last equality for all
$(k,\ell)\in R_{N}$ is easily deduced from the results in Section 5.5
of~\cite{jain89}; more details can be seen in~\cite{Kay88}.
\end{remark}

Finally we define $\zeta=(\zeta(i,j))_{0\leq i\leq N-1,0\leq j\leq N-1}$ by
\[
\zeta(k,\ell)=(\theta\ast\xi)(k,\ell)=N{\mathcal F}^{-1}((\psi{\mathcal F}%
(\xi)))(k,\ell),
\]
where $\xi=(\xi(k,\ell))_{(k,\ell).\in R_{N}}$ is a Gaussian white noise with
standard deviation $1$.

Then it is easy to prove that $\zeta=(\zeta(i,j))_{0\leq i\leq N-1,0\leq j\leq
N-1}$ is a stochastic process such that $\zeta(i,j)$ is a standard Gaussian
random variable with correlation function $\tau_{\zeta}$ satisfying~(\ref{dos}).

\subsection{Implementation}

The results presented in previous sections were implemented using the
IDL~Version 5.3 Win 32~\cite{idl521} development platform, with the following algorithm:

\begin{algorithm}
Input: $\alpha<-1$, $\gamma>0$, $n\geq1$ integer, $\rho$ and $\tau$ functions as above, then:
\begin{enumerate}
\item Compute the frequency domain mask $\psi(k,\ell)=\sqrt{{\mathcal F}%
(\tau)(k,\ell)}$.

\item Generate $\xi=(\xi (k,\ell))_{(k,\ell)\in R_{N}}$, the Gaussian white noise with zero mean and variance $1$.

\item Calculate $\zeta(k,\ell)=N{\mathcal F}^{-1}((\psi\cdot{\mathcal F}%
(\xi)))(k,\ell)$, for every $(k,\ell)$.

\item Obtain $Z_{A}^{1}(k,\ell)=G^{-1}(\Phi(\zeta(k,\ell)),(\alpha,1,n))$, for every $(k,\ell)$.

\item Return $Z_{A}(k,\ell)=\sqrt{\gamma}Z_{A}^{1}(k,\ell)$ for every
$(k,\ell)$.
\end{enumerate}
\end{algorithm}

\section{Simulation results}

In practice both parametric and non-parametric correlation structures are of interest.
The former rely on analytic forms for $\rho$, while the latter
merely specify values for the correlation.
Parametric forms for the correlation structure are simpler to specify, and its inference amounts to estimating a few numerical values; non-parametric forms do not suffer from lack of adequacy, but demand the specification (and possibly the estimation) of potentially large sets of parameters.

In the following examples the technique presented above will be used to
generate samples from both parametric and non-parametric correlation structures.

\begin{example}[Parametric situation]
\label{exampleparam}This correlation model is very popular in applications.
Consider $L\geq2$ an even integer, $0<a<1$, $0<\varepsilon
$ (for example $\varepsilon=0.001$), $\alpha<-1$ and $n\geq1$. Let
$h\colon \mathbb{R\longrightarrow R}$ be defined by
\[
h(x)=\left\{
\begin{array}
[c]{ccc}%
x & \mathrm{if} & \left|  x\right|  \geq\varepsilon,\\
0 & \mathrm{if} & \left|  x\right|  <\varepsilon.
\end{array}
\right.
\]
Let $\rho\colon R_{1}\longrightarrow(-1,1)$ be defined by $\rho(0,0)=1$ if $(k,\ell
)\neq$ $(0,0)$ in $R_{1}$ by:
\[
\rho(k,\ell)=\left\{
\begin{array}[c]{ccc}%
h(a\exp(-k^{2}/L^{2})) & \mathrm{if} & k\geq\ell,\\
-h(a\exp(-\ell^{2}/L^{2})) & \mathrm{if} & k<\ell.
\end{array}
\right.
\]
The image shown in Figure~\ref{simulated1}, of size $128\times128,$ was
obtained assuming $a=0.4$, $L=2$, $\alpha=-1.5$, $\gamma=1.0\ $and $n=1$.

\begin{figure}[h]
\begin{center}
\includegraphics[width=15cm]{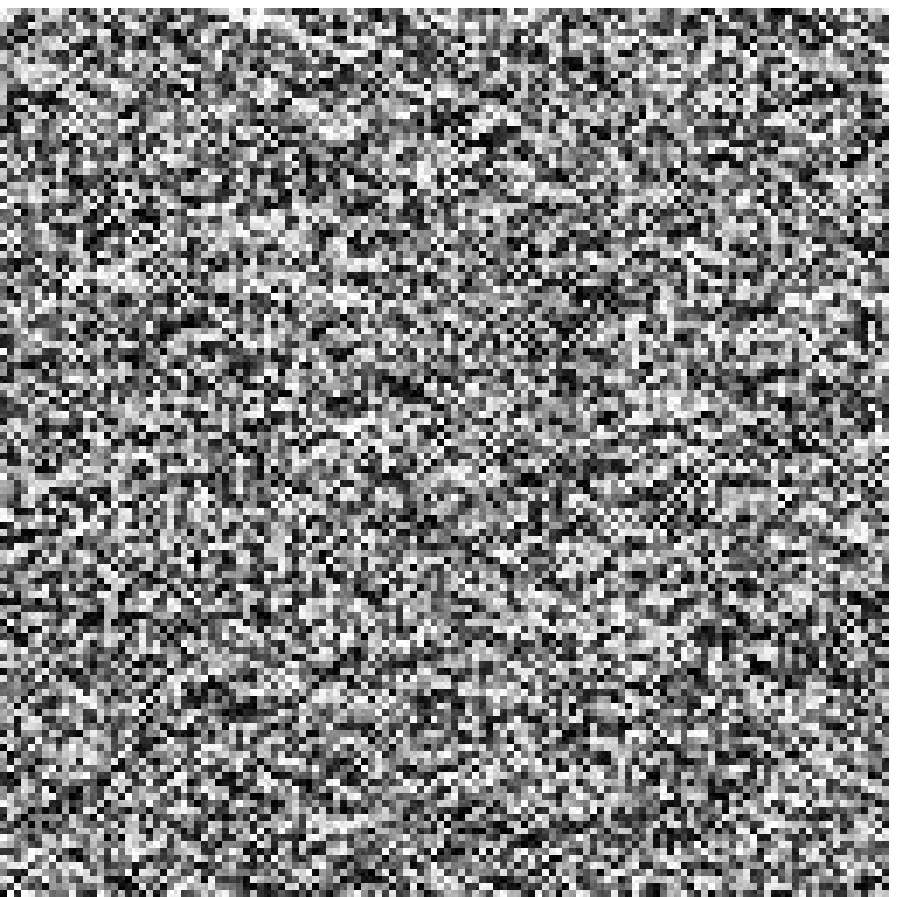}
\caption{Correlated ${\mathcal G}^{0}(-1.5,1,1)$-distributed amplitude image
with the correlation structure defined in Example~\ref{exampleparam}.}%
\label{simulated1}%
\end{center}
\end{figure}
\end{example}

\begin{example}[Mosaic] A mosaic of nine simulated fields is shown in Figure~\ref{mosaico3}.
Each field is of size $128\times128$ and obeys the model presented in Example~\ref{exampleparam}
with $a=0.4$, $\gamma =1.0$, $n=1$, roughness $\alpha$ varying in the rows ($-1.5$, $-3.0$ and $-9.0$ from top to bottom) and correlation length $L$ varying along the columns ($2$, $4$ and $8$ from left to right).
\begin{figure}[h]
\begin{center}
\includegraphics[width=15cm]{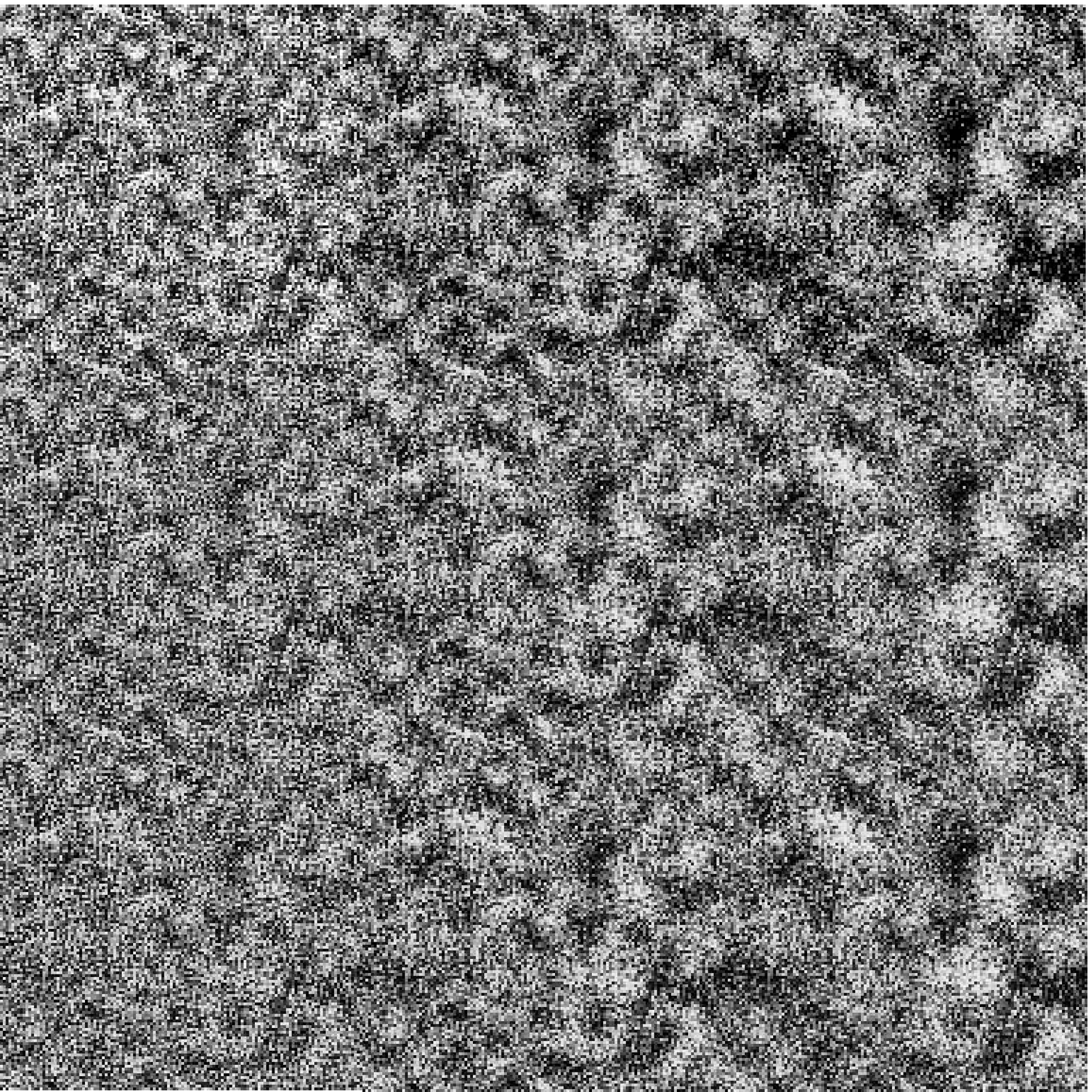}
\caption{Mosaic of nine simulated fields.}
\label{mosaico3}
\end{center}
\end{figure}
\end{example}

\begin{example}[Non-parametric situation]
\label{examplenonparam}The starting point is the urban area seen in
Figure~\ref{imaurbanreal}.
This $128\times128$ pixels image is a small sample of data obtained by the E-SAR system over an urban area.
The complete dataset was used as input for estimating the correlation structure defined by an $16\times16$ correlation matrix using Pearson's procedure ($\hat{\rho}$ below, where only values bigger than $10^{-3}$ are shown; see appendix~\ref{app_pearson}).
The correlation structure for the Gaussian process is $\tau$ below, where only values bigger than $10^{-3}$ are shown.
The roughness and scale parameters were estimated using the moments technique.
The simulated ${\mathcal G}_{A}^{0}$ field is shown in Figure~\ref{imaurbansim}.
\[
\hat{\rho}=\left(
\begin{array}
[c]{ccc}%
1.00 & 0.65 & 0.22 \\
0.97 & 0.63 & 0.22\\
0.88 & 0.58 & 0.21\\
0.76 & 0.50 & 0.19\\
0.64 & 0.43 & 0.16\\
0.53 & 0.36 & 0.14\\
0.43 & 0.30 & 0.12\\
0.36 & 0.25 & 0.10\\
0.29 & 0.20 & 0.00\\
0.24 & 0.17 & 0.00\\
0.20 & 0.13 & 0.00\\
0.16 & 0.11 & 0.00\\
0.13 & 0.00 & 0.00\\
0.11 & 0.00 & 0.00
\end{array}
\right)  ,\tau=\left(
\begin{array}
[c]{ccc}%
1.00 & 0.76 & 0.32\\
0.98 & 0.74 & 0.32\\
0.93 & 0.70 & 0.31\\
0.85 & 0.63 & 0.28\\
0.75 & 0.56 & 0.24\\
0.68 & 0.49 & 0.21\\
0.56 & 0.42 & 0.18\\
0.49 & 0.36 & 0.16\\
0.41 & 0.294 & 0.00\\
0.35 & 0.25 & 0.00\\
0.29 & 0.20 & 0.00\\
0.24 & 0.17 & 0.00\\
0.20 & 0.00 & 0.00\\
0.17 & 0.00 & 0.00
\end{array}
\right)
\]

\begin{figure}[h]
\begin{center}
\includegraphics[width=15cm]{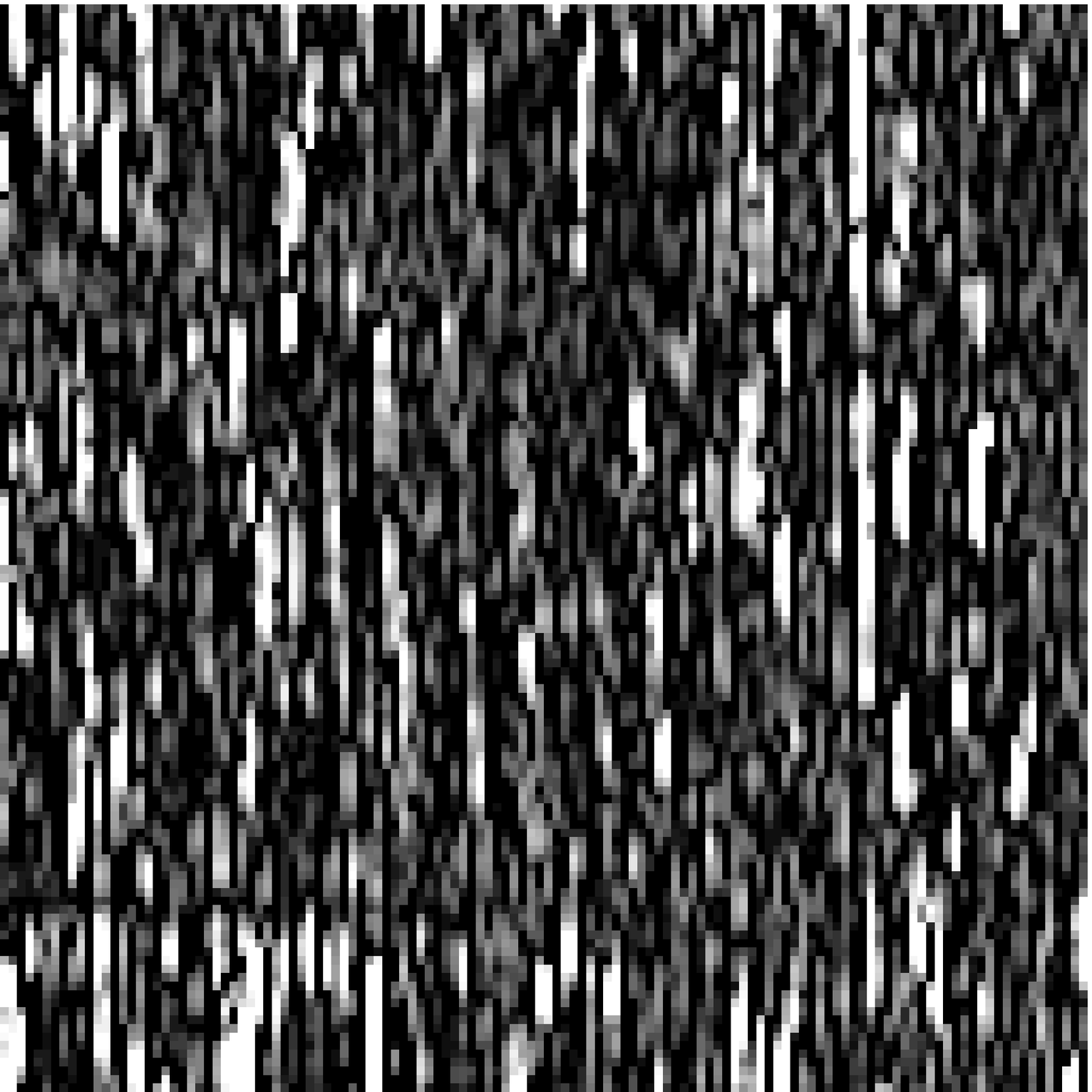}
\caption{Urban area as seen by the E-SAR system.}
\label{imaurbanreal}
\end{center}
\end{figure}

\begin{figure}[h]
\begin{center}
\includegraphics[width=15cm]{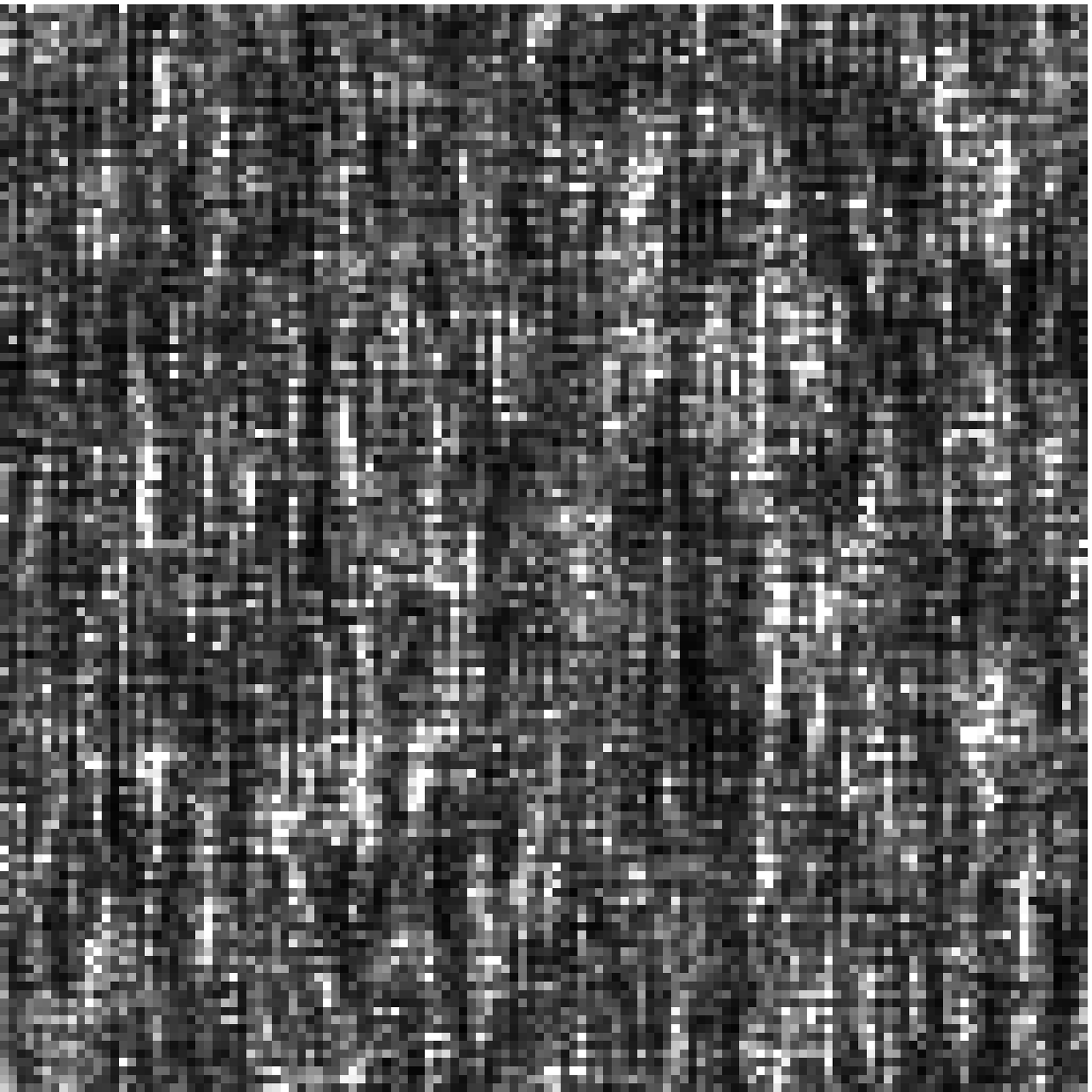}
\caption{Simulated urban area using a non-parametric correlation structure.}
\label{imaurbansim}
\end{center}
\end{figure}
\end{example}

\section{Conclusions and future work}

A method for the simulation of correlated clutter with desirable marginal law
and correlation structure was presented. This method allows the obtainment of
precise and controlled first and second order statistics, and can be easily
implemented using standard numerical tools.

The adequacy of the method for the simulation of several scenarios will be
assessed using real data, following the procedure presented in
Example~\ref{examplenonparam}: estimating the underlying correlation structure
and then simulating fields with it. A mosaic of true and synthetic textures
will be composed and made available for use in algorithm assessment.

\section*{Acknowledgements}
This work was partially supported by Conicor and SeCyT (Argentina) and CNPq (Brazil).

\appendix

\section{Estimating correlation structure with Pearson's
method\label{app_pearson}}

Consider the image $\mathbf{z}$ with $M$ rows and $N$ columns%
\[
\mathbf{z}=\left[
\begin{array}
[c]{ccc}%
z(0,0) & \cdots & z(N-1,0)\\
\vdots & \ddots & \vdots\\
z(0,M-1) & \cdots & z(N-1,M-1)
\end{array}
\right]
\]
and $n_{v}$ a positive integer smaller than $\min(M,N)$. Define $n_{c}=\left[
N/(2n_{v})\right]  $ and $n_{f}=\left[  M/(2n_{v})\right]  $, where $\left[
x\right]  =\max\left\{  k\in\mathbb{N}\colon k\leq x\right\}  $ for every real
number $x$. For each $i=0,\ldots,n_{c}-1$ and each $j=0,\ldots,n_{f}-1$ define
$\mathbf{c}(i,j)$ the submatrix of $\mathbf{z}$ of size $2n_{v}\times2n_{v}$
given by%
\[
\mathbf{c}(i,j)=\left[
\begin{array}
[c]{ccc}%
z(2n_{v}i,2n_{v}j) & \cdots & z(2n_{v}i+2n_{v}-1,2n_{v}j)\\
\vdots & \ddots & \vdots\\
z(2n_{v}i,2n_{v}j+2n_{v}-1) & \cdots & z(2n_{v}i+2n_{v}-1,2n_{v}j+2n_{v}-1)
\end{array}
\right]  ,
\]
and let $\mathbf{z}_{v}(i,j)$ be the submatrix of $\mathbf{c}(i,j)$ of size
$n_{v}\times n_{v}$ given by%
\[
\mathbf{z}_{v}(i,j)=\left[
\begin{array}
[c]{ccc}%
z(2n_{v}i,2n_{v}j) & \cdots & z(2n_{v}i+n_{v}-1,2n_{v}j)\\
\vdots & \ddots & \vdots\\
z(2n_{v}i,2n_{v}j+n_{v}-1) & \cdots & z(2n_{v}i+n_{v}-1,2n_{v}j+n_{v}-1)
\end{array}
\right]  .
\]

We will consider that $\mathbf{z}_{v}(i,j)$, for every $i=0,\ldots,n_{c}-1$
and every $j=0,\ldots,n_{f}-1$ is a sample of the random matrix%
\[
\mathbf{Z}=\left[
\begin{array}[c]{ccc}
Z(0,0) & \cdots & Z(n_{v}-1,0)\\
\vdots & \ddots & \vdots\\
Z(0,n_{v}-1) & \cdots & Z(n_{v}-1,n_{v}-1)
\end{array}
\right]  .
\]

The autocorrelation function of the random matrix $\mathbf{Z}$ is defined as
\[
\rho_{\mathbf{Z}}((m,n),(k,\ell))=\frac{E(Z(m,n)Z(k,\ell))-\mu_{Z}(m,n)\mu
_{Z}(k,\ell)}{\sigma_{Z}(m,n)\sigma_{Z}(k,\ell)},
\]
where $\mu_{Z}(k,\ell)=E(Z(k,\ell))$ and $\sigma_{Z}(k,\ell)=\sqrt
{Var(Z(k,\ell))}$, for every $0\leq m,n,k,\ell\leq n_{v}-1$.

The function $\rho_{\mathbf{Z}}$ can be estimated using Pearson's sample
correlation coefficient based on $\mathbf{z}_{v}(i,j)$, $i=0,\ldots,n_{c}-1$
and $j=0,\ldots,n_{f}-1$, i.e., for $0\leq m,n,k,\ell\leq n_{v}-1$ by%
\[
r_{\mathbf{Z}}((m,n),(k,\ell))=\frac{C_{\mathbf{Z}}((m,n),(k,\ell
))}{s_{\mathbf{Z}}(m,n)s_{\mathbf{Z}}(k,\ell)},
\]
where%
\begin{align*}
C_{\mathbf{Z}}((m,n),(k,\ell))  &  =\sum\limits_{j=0}^{n_{f}-1}\sum
\limits_{i=0}^{n_{c}-1}\left(  z(2n_{v}i+m,2n_{v}j+n)-\overline{z}%
(m,n)\right)  \left(  z(2n_{v}i+k,2n_{v}j+\ell)-\overline{z}(k,\mathbf{\ell
})\right)  ,\\
s_{\mathbf{Z}}(m,n)  &  =\sqrt{\sum\limits_{j=0}^{n_{f}-1}\sum\limits_{i=0}%
^{n_{c}-1}\left(  z(2n_{v}i+m,2n_{v}j+n)-\overline{z}(m,n)\right)  ^{2}},\\
\overline{z}(m,n)  &  =\frac{1}{n_{c}n_{f}}\sum\limits_{j=0}^{n_{f}-1}%
\sum\limits_{i=0}^{n_{c}-1}z(2n_{v}i+m,2n_{v}j+n).
\end{align*}

\end{document}